\documentclass[12]{article}
\usepackage{graphics,graphicx,amsbsy,amssymb,color}

\newcommand{\JJ}{{\boldmath \mbox{$J$}}}

\newcommand{\aaa}{{\boldmath \mbox{$a$}}}

\newcommand{\uu}{{\boldmath \mbox{$u$}}}

\newcommand{\rr}{{\boldmath \mbox{$r$}}}

\newcommand{\vv}{{\boldmath \mbox{$v$}}}

\newcommand{\nn}{{\boldmath \mbox{$n$}}}

\newlength{\defbaselineskip}
\newcommand{\setlinespacing}[1]%
           {\setlength{\baselineskip}{#1 \defbaselineskip}}

\title{\textbf{
Roles of Energy and Entropy in Multiscale Dynamics and Thermodynamics}}
\author{Miroslav Grmela \footnote{ e-mail:
miroslav.grmela@polymtl.ca}\\
\'{E}cole Polytechnique de Montr\'{e}al,
  C.P.6079 suc. Centre-ville,\\
 Montr\'{e}al, H3C 3A7,  Qu\'{e}bec, Canada}

 \date{}

\begin{document}

\maketitle

\tableofcontents
\setlength{\parskip}{4mm}

\begin{abstract}

Multiscale thermodynamics is a theory of relations among  levels of description. Energy and entropy are its two main ingredients.
Their roles in the time evolution describing  approach of a level (starting level) to another level involving less details (target level) is examined on several examples, including the level on which macroscopic systems are seen as composed of microscopic particles,  mesoscopic levels as kinetic theory of ideal and van der Waals gases, fluid mechanics, the level of chemical kinetics, and the level of equilibrium thermodynamics. The entropy enters the emergence of the target level in two roles.  It expresses  internal energy, that is the part of the energy that cannot be expressed in terms of the  state variables used on the starting level, and it  reveals  emerging features characterizing the target  level by  sweeping away unimportant details. In the case when the target level is a mesoscopic level involving  time evolution the  roles of the energy and the entropy is taken  by  two different potentials that are related to their  rates.

\end{abstract}

\section{Introduction}

Our point of departure is a pair of autonomous levels of description of macroscopic systems. Both have arisen from certain type of experimental observations. The types of observations are different different  levels. One level, that we call  \textit{starting level}, is based on more detailed observations than  the second level called  \textit{target level}. We emphasize  that both levels are autonomous in the sense that both levels separately provide a good description of the experimentally observed behavior without a need of other levels. In particular, we  investigate the levels of particle mechanics, kinetic theory and fluid mechanics as starting levels. All these  three levels are paired with the level of the classical equilibrium thermodynamics as the target level. In Section \ref{Rate} we  also investigate pairs of levels in which both the starting and the target levels are  mesoscopic levels involving  time evolution.

Since we assume that both the starting and the target levels are autonomous and well established (i.e. predictions of the theory agree with results of observations), one observation (among all observations made on the starting level) has to be an observation of the approach to the target level. In other words, the autonomous existence of both levels guarantees  the  possibility  to prepare macroscopic systems for the target level and the visibility of the preparation process in the time evolution observed  on the starting level. We emphasize that in passing from the starting level to a target level that involves less details we loose  details but gain  emerging overall features.

We use  the following terminology.
Solutions of the governing equations on the starting level are trajectories, their collection is called a \textit{phase portrait}. We distinguish  three types of energy. Sources of the \textit{external energy} are outside the macroscopic system under investigation. The potential energy in  gravitational field is an example of the external energy. The \textit{inner energy} is the energy that can be expressed in terms of  the state variables used on the  chosen level of description. The total energy in the Gibbs theory or the kinetic energy of a fluid in fluid mechanics are examples of the inner energy. The former is expressed in  terms of the $n$-particle distribution function that serves as the state variable in the Gibbs theory and the latter in terms of the velocity and mass fields that serve as state variables in fluid mechanics. The \textit{internal energy} is the energy that cannot be expressed in terms of the state variables that are used on the chosen level of description.  Microscopic details  that are not seen on the chosen level of description are needed to express it.

The objective of this paper is to investigate the roles that  the energy and that entropy  play in the passage from the starting level to the target level.

\textit{\textbf{Energy} provides  the force generating the phase portrait.}

\textit{\textbf{Entropy} plays two roles:}
\textit{(Ent I) it expresses the internal energy, }
\textit{(Ent II) it  makes  patterns, emerging in the phase portrait  during the time evolution,  manifestly visible by  sweeping  away unimportant details.  The patterns represent the target levels inside the starting level.}

We explore this viewpoint of energy and entropy in the setting of  GENERIC time evolution  (a combination of  Hamiltonian and gradient dynamics) describing the passage from a starting level to a target level. The "universal competition"  between   energy and entropy is also discussed, but with a different perspective, in \cite{MW}.

Before starting our  discussion
we briefly recall history of GENERIC dynamics.
The first step in its formulations,  made  by Vladimir Arnold  \cite{Arnold}, was  casting  the  Euler fluid mechanics  into the form of noncanonical Hamiltonian dynamics.
Vladimir Arnold  has also realized  connection  with  older results obtained in  investigations of Lie groups. Such  connection then allowed to formulate  nondissipative parts of  kinetic  \cite{MorrPRL} and other mesoscopic time evolution equations as noncanonical Hamilton's equations. In proceedings of the conference devoted to this subject  (organized by Jerrold  Marsden in the summer of 1983 in Boulder, Colorado)   the formulation of the  complete Boltzmann kinetic equation (that includes the dissipative collision term) as a combination of Hamiltonian and generalized gradient dynamics has appeared \cite{GrmelaBE}. Many other mesoscopic  time evolution equations have been put into such  form in \cite{MorrPla}, \cite{Kaufman}, \cite{GrmelaPla}, \cite{MorrPhys}, \cite{GrmelaPhys}, \cite{GEN2}. In \cite{GO}, \cite{OG} the combination of  Hamiltonian and  generalized gradient dynamics has been called GENERIC (an acronym for General Equation for Non Equilibrium Reversible Irreversible Coupling). The natural geometrical setting for the Hamilton-gradient dynamics is contact geometry \cite{Grmelacont}, \cite{EssenGP}.

\section{\textit{starting level}$\longrightarrow$ \textit{level of equilibrium thermodynamics}}\label{sec1}

The target level in this section is the level of the classical thermodynamics, the starting levels vary. Historically, the  passages of this type were  investigated by Boltzmann \cite{Boltzmann},  Gibbs \cite{Gibbs}, and  Prigogine \cite{Prigogine}. Their comparison leads to the abstract formulation presented below. Its four particular realizations (Boltzmann's kinetic theory, Gibbs equilibrium statistical mechanics, Navier-Stokes-Fourier fluid mechanics,  van der Waals theory)  provide an insight into the roles of energy and entropy.

\subsection{Static theory (MaxEnt)}\label{sfgen}

The state variable chosen on the starting level is denoted by the symbol $x$, the energy is $E(x)$, the entropy $S(x)$, and the number of moles  $N(x)$. In the Gibbs theory $x$ is the $n$-particle distribution function ($n\sim 10^{23}$), in the Boltzmann theory, discussed in Section \ref{B},  the state variable $x$ is the one particle distribution function, in fluid mechanics theories discussed in Section \ref{FM} the state variable $x$ is a collection of  hydrodynamic fields, and in the classical equilibrium thermodynamics the state variable $x=(E,N,V)$, where $E$ is the energy, $N$ number of moles, and $V$ is the volume.

The input into the passage \textit{starting level}$\longrightarrow$ \textit{level of equilibrium thermodynamics}  is the \textit{fundamental thermodynamic relation on the starting level} consisting of  three real valued functions
\begin{equation}\label{xfre}
S=S(x);\,\,y=y(x)
\end{equation}
where $y=\left(\begin{array}{cc}E\\N\end{array}\right)$  are state variables on the target level that is in this section always the level of the classical equilibrium thermodynamics, $E$ is the equilibrium thermodynamic energy and $N$ is the equilibrium thermodynamic number of moles. By $y(x)=\left(\begin{array}{cc}E(x)\\N(x)\end{array}\right)$ we denote the energy and the number of moles given in the fundamental thermodynamic relation (\ref{xfre}) on the starting level.
Both  functions  $S(x)$ and $y(x)$ are assumed to be sufficiently regular and the entropy $S(x)$ is moreover assumed to be concave.

With (\ref{xfre}) we construct thermodynamic potential
\begin{equation}\label{Phi}
\Phi(x,y^*)=-S(x)+<y^*,y(x)>
\end{equation}
The covector   $y^*=(E^*,N^*)$  denotes the Lagrange multipliers. We require
\begin{equation}\label{estar}
E^*>0
\end{equation}
In the standard equilibrium thermodynamic notation $E^*=\frac{1}{T}$ and $N^*=-\frac{\mu}{T}$, where $T$ is the absolute equilibrium thermodynamic temperature in energy units and $\mu$ is the equilibrium thermodynamic chemical potential. The requirement (\ref{estar}) is thus the requirement that the absolute temperature $T$ is positive. By $<,>$ we denote pairing in the equilibrium thermodynamic state space; $<y^*,y>=E^*E+N^*N$.

With the thermodynamic potential $\Phi(x,y^*)$  we are now in position to make
the passage $staring\,\,level\,\,\rightarrow\,\,target\,\,level$.  The passage is made by
\begin{equation}\label{maxent}
minimizing\,\,the\,\,thermodynamic\,\,potential\,\,\Phi(x,y^*)
\end{equation}

Solutions of
\begin{equation}\label{Phix}
\Phi_x(x,y^*)=0
\end{equation}
is denoted $x_{eq}(y^*)$ and called equilibrium states. We use a shorthand notation $\Phi_x=\frac{\partial\Phi}{\partial x}$. If $x$ is an element of an infinite dimensional space (e.g. a distribution function) then the derivative is an appropriate functional derivative.

The Legendre transformation $S^*(y^*)$ (that belongs to the target level)  of $S(x)$ (that belongs to the starting level) is given by
\begin{equation}\label{Sstar}
S^*(y^*)=\Phi(x,y^*)|_{x=x_{eq}(y^*)}
\end{equation}
The function $S^*(y^*)$ is a Legendre transformation of $S(y)$ that is the entropy on the target level  implied by the entropy $S(x)$ on the starting level.  Explicitly,
\begin{equation}\label{Sy}
S(y)=\Phi^*(y^*,y)_{y^*=y^*_{ex}(y)}
\end{equation}
where $\Phi^*(y^*,y)=-S^*(y^*)+<y^*,y>$ and $y^*_{ex}(y)$ is a solution of $\Phi^*_{y^*}(y^*,y)=0$.

So far we have not addressed yet  the  region $\Omega\subset \mathbb{R}^3$ in which the macroscopic system under investigation is confined. We do it now but only on the target level. We  characterize $\Omega$ only by its volume  $V$. We take it into account by extending
 $S(y)$ introduced  in (\ref{Sy}) into $S=S(y,V)$. We  require  that both $S$ and $y$ are extensive state variables in the sense  that $S(\lambda E,\lambda N,\lambda V)=\lambda S$, where $\lambda\in\mathbb{R}$. From the Euler relation we have $S=<\frac{\partial S}{\partial y}, y>+
\frac{\partial S}{\partial V} V$. From (\ref{Sy}) we have $\frac{\partial S}{\partial y}=y^*$. The Euler relation takes the form $S=<y^*,y>+V^*V$, where $V^*=\frac{\partial S}{\partial V}$. In the standard notation of the classical thermodynamics $V^*=\frac{P}{T}$, where $P$ is the equilibrium pressure. Consequently,
\begin{equation}\label{SstarV}
S^*(y^*)=-V^*
\end{equation}

Summing up, we have passed from the starting level with state variables $x$ and the fundamental thermodynamic relation (\ref{xfre}) to the target level with  state variables $y$. The input is the fundamental thermodynamic relation (\ref{xfre}), the output  is the equilibrium state $f_{eq}(y^*)$ and the fundamental thermodynamic relation
\begin{equation}\label{yfre}
S=S(y);\,\,y=y
\end{equation}
The equilibrium state  $f_{eq}(y^*)$ places the  level of equilibrium thermodynamics inside the starting level and $S(y)$ is the fundamental thermodynamic relation on the  level of equilibrium thermodynamics that is inherited from the fundamental thermodynamics relation on the starting level.
We note that the passage $starting\,\,level\,\,\rightarrow\,\,target\,\,level$ made by MaxEnt (\ref{maxent}) is in fact a reducing Legendre transformation.

The conjugate variables $y^*$ introduced in the thermodynamic potential (\ref{Phi}) belong to the target level (that is in this section the level of equilibrium thermodynamics).
One of them, namely $e^*=\frac{1}{T}$, relates  energy to  entropy and is therefore of particular interest in this paper. We shall discuss it in more detail in Section \ref{eegen} below.

\subsection{Dynamic theory}\label{dfgen}

The postulated maximization of the entropy (\ref{maxent}) is
replaced in the dynamic formulation by the time evolution equation
\begin{equation}\label{generic}
\frac{\partial x}{\partial t}=L(x)\mathfrak{E}_x(x)-\Xi_{x^*}(x,x^*)|_{x^*=\mathfrak{E}_x}
\end{equation}
where
\begin{equation}\label{avE}
\mathfrak{E}(x,y^*)=\frac{1}{e^*}\Phi(x,y^*)=E(x)-\frac{1}{e^*}S(x)+\frac{n^*}{e^*}N(x)
\end{equation}
is called, in accordance with the standard terminology (see (\cite{MW}), an available free energy.
The maximization of the entropy $S(x)$ subjected to constraints $E(x),N(x)$ is made by following the time evolution governed by  (\ref{generic}). The asymptotic solution $t\rightarrow\infty$ of (\ref{generic}) is the equilibrium state $x_{eq}(y^*)$ that is also a solution to (\ref{Phix}).

We now explain  the meaning of the symbols introduced in the GENERIC equation (\ref{generic}) and show that its solutions indeed approach $x_{eq}(y^*)$. The first term on the right hand side of (\ref{generic}) is the Hamiltonian vector field. It  is the covector $\mathfrak{E}_x$ transformed into vector by the Poisson bivector $L(x)$. This bivector is defined by the bracket $\{A,B\}=<A_x,LB_x>$ that is required to be   a Poisson bracket satisfying the following properties:
\begin{eqnarray}\label{L}
&&antisymmetry\,\,\{A,B\}=-\{B,A\}\nonumber \\
&&Jacobi\,\, identity\,\,\{A,\{B,C\}\}+\{B,\{C,A\}\}+\{C,\{A,B\}\}=0
\end{eqnarray}
$A(x)$ and $B(x)$ are real valued and sufficiently regular function of $x$.
From the physical point of view, the Poisson bracket expresses the kinematics of the state variable $x$.

Next, we  turn to  the second term on the right hand side of (\ref{generic}).
The symbol $\Xi$ stands for a real valued function of $(x,x^*)$, called a dissipation potential,  satisfying the following properties
\begin{eqnarray}\label{Xi}
&&\Xi(x,x^*)|_{x^*=0}=0\nonumber \\
&&\Xi(x,x^*)\,\,reaches\,\,its\,\,minimum\,\,at\,\,x^*=0\nonumber \\
&&\Xi(x,x^*)\,\,is\,\,a\,\,convex\,\,function\,\,of\,\,x^*\,\,in\,\,a\,\,neighborhood\,\,of\,\,x^*=0\nonumber \\
\end{eqnarray}
We note that for small $x^*$ (i.e. in a small neighborhood of equilibrium states) all dissipation potentials are quadratic functions of $x^*$.

With the requirements (\ref{L}), (\ref{Xi}), (\ref{deg}) Eq.(\ref{generic}) implies
\begin{equation}\label{Liap}
\frac{\partial \Phi}{\partial t}\leq 0
\end{equation}
This inequality  makes the thermodynamic potential (\ref{Phi}) the Lyapunov function  (provided (\ref{estar}) is taken into account and provided not only   $-S(x)$ but also $\Phi(x)$ is a convex function) indicating the approach to $x_{eq}(y^*)$. A rigorous proof of  the approach requires an additional analysis. We shall comment about it at the end of Section \ref{B}.

If we equip (\ref{generic}) with an extra structure
\begin{eqnarray}\label{deg}
&&LS_x=0;\,\,LN_x=0\nonumber \\
&&\Xi\,\,depends\,\,on\,\,x^*\,\,only\,\,through\,\,the\,\,dependence\,\,on\nonumber\\
&& the\,\,thermodynamic\,\,force\,\,X=Kx^*\,\,
where\,\,K\,\,is\,\,a\,\,linear\,\,operator\nonumber \\
&&satisfying\,\,KE_x=0;\,\,KN_x=0
\end{eqnarray}
then (\ref{generic}) turns into the familiar GENERIC equation ( see e.g. \cite{book})
\begin{equation}\label{generic1}
\frac{\partial x}{\partial t}=L(x)E_x(x)+\widehat{\Xi}_{x^*}(x,x^*)|_{x^*=S_x}
\end{equation}
where $\widehat{\Xi}=\frac{1}{e^*}\Xi$. Equation (\ref{generic1})  with the structure (\ref{deg} then implies
\begin{eqnarray}\label{prop}
\frac{dE}{dt}&=&0\nonumber\\
\frac{dN}{dt}&=&0\nonumber \\
\frac{dS}{dt} &\geq&0
\end{eqnarray}
is a stronger property than (\ref{Liap}). The result (\ref{prop}) implies (\ref{Liap}) but  gives a more information about solutions to (\ref{generic1}).

\subsection{Roles of Energy and Entropy}\label{eegen}

The roles that the energy and the entropy play on various  starting levels depend on the  levels. They  will be discussed below. Here we recall the roles that they play on the target level that is in this section the level of the classical equilibrium thermodynamics.
On this level the entropy plays the role of one of the  state variables. There is no inner energy, the only energy is the internal energy. Moreover, the internal energy  is in  one-to-one relation with the entropy. The entropy thus plays only the role (\textit{Ent I}).

There is no time evolution on the level of equilibrium thermodynamics. The entropy therefore does not play the role \textit{(Ent II)}. However, equilibrium thermodynamics considers  processes in which states  change.  The  changes are  due to interactions through walls surrounding the macroscopic systems. The walls can pass or prevent passing the internal energy, can expand or shrink the system, and can enlarge or diminish the number of moles. An investigation of the time evolution involved in such changes would require to step outside the level of equilibrium thermodynamics on a level involving more details. When remaining inside the equilibrium level, the processes  are  considered  as sequences of equilibrium states. Their  final outcome  is determined  by MaxEnt. In this sense the entropy plays on the level of equilibrium thermodynamics also the role \textit{(Ent II)}. The unimportant details that are swept away in this pattern recognition process are details of initial arrangements of the subsystems.

Since the relation between internal energy and entropy is invertible their roles as state variables can be exchanged. The fundamental thermodynamic relation $E=E(S,N,V)$  can be replaced by
\begin{equation}\label{eqfund}
S=S(E,N,V)
\end{equation}
 In addition,
the set of state variables $(S,N,V)$ can be enlarged by adopting  variables characterizing  overall states of  macroscopic systems. For example it can be the  overall velocity or coordinates characterizing placement of the macroscopic system in  an imposed force field (e.g. gravitational field). In such case the total energy is a sum of the inner energy (the energy that can be expressed in terms of the newly adopted state variables, for example the  overall kinetic and/or potential energy) and the remaining internal energy  expressed in terms of $(S,N,V)$.

Because our objective in this paper is to investigate relations between energy and entropy, a particularly  important  concept is the temperature. On the level of equilibrium thermodynamics the temperature is defined by $\frac{1}{T}=S_E$ (see (\ref{Sy})). MaxEnt principle  implies that two systems, that are connected by a wall that freely passes the internal energy $E$ and both are surrounded by a wall that prevents such passing approach an equilibrium state at which both systems,  have the same temperature. This is the way the temperature on the level of equilibrium thermodynamics is measured. One of the two systems is a thermometer for which the fundamental thermodynamic relation is known and thus its temperature can be  read in other quantities (e.g. volume).
A general temperature can be defined \cite{mt} as a measure of  internal energy, its measurements involve a process of equilibration. Different meanings that can be given to   "internal energy" and  "equilibration" lead to different meanings of the temperature.

\section{Particular Realizations}\label{PR}

The multiscale formulation of the passage \textit{starting level}$\rightarrow$ \textit{level of equilibrium thermodynamics} (Section \ref{sec1})) is a common  structure extracted from investigations of many such passages with specific choices of the starting level. Some of these passages are now presented as particular realizations of the multiscale formulation. In all realizations we  always begin with the static theory, continue with the dynamic theory and end with  a discussion of the roles of energy and entropy.

\subsection{Boltzmann's kinetic theory}\label{B}

Boltzmann's investigation \cite{Boltzmann} of dynamics of ideal gases was the first step towards understanding the approach of macroscopic systems to equilibrium.  The insight  allowing to recognize the equilibrium pattern in the phase portrait is the realization that collisions of  gas particles are the source of unimportant details that have to be swept away in order that the equilibrium pattern is revealed.  We present below the Boltzmann theory as a particular realization of the multiscale formulation of \textit{starting level}$\rightarrow$ \textit{level of equilibrium thermodynamics}.

Before placing ourselves on the level of kinetic theory,  we recall that on the level of equilibrium thermodynamics the individual nature of an ideal gas is expressed in the  equilibrium-thermodynamics fundamental thermodynamic relation
\begin{equation}\label{eqideal}
S=\frac{5}{2}k_BN+NR\ln \left[\left(\frac{E}{V}\right)^{3/2}\left(\frac{V}{N}\right)\right]
\end{equation}
where $k_B$ is the Boltzmann constant and $R$ the universal gas constant. On the level of equilibrium thermodynamics, this relation is   obtained from  experimental observations of the behavior of ideal gases. Our objective is to get (\ref{eqideal}) from  investigating the passage \textit{level of kinetic theory}$\rightarrow$\textit{level of equilibrium thermodynamics}. Our objective is to introduce a particular realization of  (\ref{generic}) which implies (\ref{eqideal}).

We begin with the static theory.
The state variable on the level of kinetic theory is the one particle distribution function
\begin{equation}\label{svkt}
x=f(\rr,\vv)
\end{equation}
 We put the mass of one particle equal to one, $\rr$ is the position coordinate and $\vv$ momentum of one particle. The ideal gas under investigation is assumed to be confined in the region $\Omega\subset \mathbb{R}^3$ with periodic boundary conditions (i.e. all integrals over the boundary that arise in by parts integrations equal zero).

The fundamental thermodynamic relation of an  ideal gas on the level of kinetic theory is
\begin{eqnarray}\label{idfund}
S(f)&=&-\int d\rr\int d\vv f(\rr,\vv)\ln f(\rr,\vv)\nonumber \\
E(f)&=&\int d\rr\int d\vv f(\rr,\vv)\frac{\vv^2}{2}\nonumber \\
N(f)&=&\int d\rr\int d\vv f(\rr,\vv)
\end{eqnarray}
The energy $E(f)$ is the kinetic energy.  The physical interpretation of $f(\rr,\vv)$ as a distribution function leads directly to the number of moles $N(f)$, and the entropy $S(f)$ is postulated.

The MaxEnt principle (\ref{maxent}) leads  (see (\ref{Phix})) to the equilibrium state
\begin{equation}\label{ideq}
f_{eq}(\rr,\vv)=\left(\frac{1}{2\pi T}\right)^{3/2}\exp \left(\frac{\mu}{T}-1\right) \exp\left(-\frac{\vv^2}{2T}\right)
\end{equation}
and the ideal gas fundamental  thermodynamic relation (\ref{eqideal}) on the level of  equilibrium thermodynamics.

Next, we turn to the dynamic theory.
The postulated MaxEnt  principle (\ref{maxent}) is replaced by the Boltzmann kinetic equation
\begin{equation}\label{BEq}
\frac{\partial f}{\partial t}=-\frac{\partial}{\partial \rr}\left(f\frac{\partial E_f}{\partial \vv}\right)-\Xi_{f^*}|_{f^*=S_f}
\end{equation}
that is a particular realization of (\ref{generic1}) with
the Poisson bracket
\begin{equation}\label{idL}
\{A,B\}= \int d\rr\int d\vv f\left(\frac{\partial A_f}{\partial \rr}\frac{\partial B_f}{\partial \vv}-\frac{\partial B_f}{\partial \rr}\frac{\partial A_f}{\partial \vv}\right)
\end{equation}
(where $A(f), B(f)$ are real valued sufficiently regular functions of $f(\rr,\vv)$)
and the dissipation potential
\begin{equation}\label{iddissp}
\Xi(f,X)=\int d\rr\int d\vv\int d\vv_1\int d\vv'\int d\vv'_1 W(f,\vv,\vv_1,\vv',\vv'_1)\left(e^X+e^{-X}-2\right)
\end{equation}
where  $W\neq 0$ only if $\vv+\vv_1=\vv'+\vv'_1$ and $\vv^2+(\vv_1)^2=(\vv')^2+(\vv'_1)^2$. In addition, $W$ is symmetric with respect to the interchange of $\vv\rightarrow \vv_1$ and $\vv'\rightarrow \vv'_1$  and  with respect to the interchange of $(\vv,\vv_1)$ and $(\vv',\vv'_1)$. The thermodynamic force $X$ is given by
\begin{equation}\label{XBol}
X(f,f^*)= \mathbb{K}f^*=f^*(\rr,\vv')+f^*(\rr,\vv'_1)- (f^*(\rr,\vv)+f^*(\rr,\vv_1)).
\end{equation}

The Poisson bracket (\ref{idL}) expresses  kinematics of the one particle distribution function $f(\rr,\vv)$ (see e.g. \cite{book}). Regarding the dissipation potential (\ref{iddissp}), a direct verification  proves that it  satisfies the required properties (\ref{Xi}).   The degeneracy requirements (\ref{deg})  can also be  proved by a direct verification.  All functions $C(f)=\int d\rr\int d\vv \zeta(f(\rr,\vv))$ where $\zeta:\mathbb{R}\rightarrow \mathbb{R}$ satisfy $\{A,C\}=0\,\, \forall A$. Such functions are called Casimirs of the bracket $\{A,B\}$. Consequently, the Poisson bivector $L$ defined by (\ref{idL}) satisfies thus the degeneracy requirement (\ref{deg}). The required degeneracy of the  dissipation potential (\ref{iddissp}) can  also be  directly  verified.

With the above specification of the building blocks of (\ref{generic}), the thermodynamic potential (\ref{Phi}) plays the role of the Lyapunov function for the time evolution governed by Eq.(\ref{BEq}). Maximization of the entropy subjected to constraints of energy and number of moles is thus made by following  the time evolution governed by  (\ref{BEq}) to its conclusion.

There are still  missing pieces in a rigorous proof of the approach of solutions of  (\ref{BEq}) to the equilibrium state (\ref{ideq}). In particular it is the existence of solutions of the Boltzmann equation and  an  additional analysis (in addition to identifying the Lyapunov function) needed to prove the approach to the equilibrium state. The former, provided in \cite{Lions}, demonstrates agreement with  experimental observations. The time evolution of ideal gases is seen in experimental observations to exist. Solutions to the Boltzmann kinetic equation describing the time evolution of ideal gases in their mathematical representation  is proven in \cite{Lions} to exist. Results proven
in \cite{GradB}, \cite{Villani1}, \cite{Villani2} are even more  physically  significant. Solutions to the Boltzmann equation with the Hamiltonian term missing approach the local equilibrium (that is the equilibrium state (\ref{ideq}) in which $n$ and $T$ are unspecified functions of the position coordinate $\rr$). It is the coupling with the Hamiltonian term  that  brings solutions to the total equilibrium (\ref{ideq}). The Hamiltonian term  by itself does not produce any dissipation.  The enhancement of dissipation,  that is due to  the  coupling of gradient dynamics with Hamiltonian dynamics,   is called Grad-Villani-Desvillettes enhancement of dissipation.   Very likely this is the principal  mechanism making  the transformation of time reversible and nondissipative Hamiltonian time evolution of $\sim 10^{23}$ particles to time irreversible and dissipative GENERIC time evolution. A very small instability on the microscopic level may be enhanced by the Grad-Villani-Desvillettes enhancement to macroscopic dissipation bringing macroscopic systems to equilibrium states.
So far only the damping that brings  the local equilibrium to the global equilibrium in the Boltzmann dynamics \cite{GradB}, \cite{Villani1}  and the Landau damping occurring in the Vlasov dynamics  \cite{Villani2}, \cite{GrPavL} have been   proven rigorously. An argument supporting the general importance of the Grad-Villani-Desvillettes enhancement of dissipation is presented in Section \ref{fmm}.

\subsubsection{Roles of Energy and Entropy}\label{eeB}

Ideal gas particles do not interact among themselves. Their kinetic energy, that is expressed in terms of $f(\rr,\vv)$ and  is thus  an inner energy,  is  the only energy.  The entropy plays only the role  \textit{(Ent II)}.  In the process of its maximization (or equivalently  in the process of following solutions of the Boltzmann kinetic equation to $t\rightarrow\infty$)  unimportant details in the phase portrait are swept away and the equilibrium pattern in the phase portrait emerges.
The entropy is not postulated in the Boltzmann theory. It arises from investigating solutions of the Boltzmann kinetic equation. The individual nature of ideal gases  is expressed only in the energy (that is the inner energy). The entropy is universal.
We also note that on the starting level there is no direct relation between the energy and the entropy. Both are functions of $f(\rr,\vv)$ but $f(\rr,\vv)$ cannot be eliminated between  them. On the other hand, on the target level the energy and the entropy are directly related, there is a one-to-one relation between them.

There is no temperature on the level of kinetic theory because there is no internal energy. Only after the inner energy has been transformed by MaxEnt (or equivalently by following the time evolution generated by the Boltzmann equation), to the internal energy on the level of equilibrium thermodynamics, the temperature can be defined   on the submanifold composed of the equilibrium states $f_{eq}(y)$ $\left(\frac{3}{2}T=(\int d\rr\int d\vv f)^{-1}\int d\rr\int d\vv \frac{\vv^2}{2}\right)$.
We can also introduce  local temperature that arises on the manifold composed of the local equilibrium states. .

Before leaving the Boltzmann theory, we note that the abstract  multiscale thermodynamics  in  Section \ref{sec1} is in fact  an extraction of the mathematical structure that is present  in the Boltzmann equation.  By transforming the original Boltzmann theory into the form presented in Section \ref{sec1} we have in fact extended its applicability to general macroscopic systems.

\subsection{Gibbs equilibrium statistical mechanics}\label{Gibbs}

.

Gibbs' investigation \cite{Gibbs} is limited to static situations but is applicable to all macroscopic systems.
The state variable is chosen to be $n$-particle distribution function $f(z)$, where  $n\sim 10^{23}$ is a fixed  number of microscopic particles composing macroscopic systems,  $z= (\rr,\vv)=(z_1,...,z_n)=(\rr_1,...,\rr_n\vv_1,...,\vv_n)$, where $z_i=(\rr_i,\vv_i)$,  is the position coordinate and the momentum if $i-th$ particle, $i=1,2,...,n$, $n\sim 10^{23}$.
The Gibbs equilibrium statistical mechanics is traditionally introduced  (see e.g. \cite{LL}) in two steps. The equilibrium distribution function $f_{eq}(z)$ in the first step, the entropy in the second step.
Gibbs assumes that $f_{eq}(z)$
depends only on constants of motion (i.e. on the energy and the number of moles).
Other details of their trajectories  do not enter  $f_{eq}(z)$  due to the ergodic hypothesis  according to which the particle trajectories are uniformly spread. The exponential  dependence of $f_{eq}(z)$  on
the energy and the number of moles then follows from noting that the energy and the number of moles   of two independent subsystems is a sum of their  energies and the numbers of moles
while $f(z)$  is a multiplication of their two distribution functions.
The Gibbs entropy appears    by requiring that $f_{eq}(z)$  arises  in MaxEnt.

Below, we present the Gibbs theory as a particular realization of the static multiscale formulation of \textit{level of particle mechanics}$\longrightarrow$ \textit{level of equilibrium thermodynamics}  in Section \ref{sfgen}. We shall also supplement the static theory with   a corresponding to it dynamic theory   that is a particular realization of the multiscale  dynamic theory  in Section \ref{dfgen}.

The state variable on the starting level of the Gibbs theory is the $n$-particle distribution function
\begin{equation}\label{svG}
x=f(z)
\end{equation}
The fundamental thermodynamic relation  is
\begin{eqnarray}\label{Gfund}
S(f)&=&-\int dz f(z)\ln f(z)\nonumber \\
E(f)&=& \int dz f(z) h(z)\nonumber \\
N(f)&=&\int dz f(z)
\end{eqnarray}
where $h(z)$ is the particle Hamiltonian  and  $k_B$ is the Boltzmann constant. With these specifications the MaxEnt principle leads to the equilibrium state
\begin{equation}\label{caneq}
f_{eq}(z)=\exp\left(\frac{\mu}{T}-1\right)\exp\left(-\frac{h(z)}{T}\right)
\end{equation}
and the fundamental thermodynamic relation.
\begin{equation}\label{canfr}
S^*(\mu,T)=\exp\left(\frac{\mu}{T}-1\right)\int dz\exp\left(-\frac{h(z)}{T}\right)
\end{equation}

Now we turn to the dynamic theory.
The Gibbs theory addresses the time evolution only in the  energy conservation and in the ergodic hypothesis.
Inspired by   Boltzmann, we  suggest a particular realization of (\ref{generic}) replacing the ergodic hypothesis by the time evolution.

The time evolution of $f(z)$  (a lift of the Hamiltonian and  reversible time evolution of $z$ to the time evolution of real valued functions of $z$ \cite{book}) is governed by the Liouville equation
\begin{equation}\label{Liouville}
\frac{\partial f(z)}{\partial t}=-\sum_{i=1}^n \sum_{\alpha=1}^3\frac{\partial}{\partial r_{i\alpha}}\left( f(z)\frac{\partial h(z)}{\partial v_{i\alpha}}\right)+\frac{\partial}{\partial v_{i\alpha}}\left( f(z)\frac{\partial h(z)}{\partial r_{i\alpha}}\right)
\end{equation}
We note that this equation is a Hamilton equation $\frac{\partial f(z)}{\partial t}=L(f)E(f)$ with the energy $E(f)$ given in (\ref{Gfund})  and the Poisson bivector $L(z)$ given by  the Poisson bracket \cite{book}
\begin{equation}\label{PBLiouville}
\{A,B\}=<A_f,L(f)B_f>=\int dz f(z)\sum_{i=1}^n\sum_{\alpha=1}^3\left(\frac{\partial A_f}{\partial r_{i\alpha}}\frac{\partial B_f}{\partial v_{i\alpha}}- \frac{\partial B_f}{\partial r_{i\alpha}}\frac{\partial A_f}{\partial v_{i\alpha}}\right)
\end{equation}
This means that the Liouville equation is a particular realization of (\ref{generic}) with the second term on its right hand side missing.
We also note that the lift from the time evolution of $z$ to the time evolution of $f(z)$ is a linearization. Typically very nonlinear equations governing the time evolution of $z$ become  linear Liouville equations (\ref{Liouville}) governing the time evolution of $f(z)$.

Can  we modify the Liouville equation (\ref{Liouville}) in such a way that: (i) it becomes a particular realization of (\ref{generic}), and (ii) the time evolution that it generates maximizes the Gibbs entropy subjected to constraints of the  energy and the number of moles (specified in (\ref{Gfund})).
Our goal  is to make the  emergence of the equilibrium pattern in the Liouville phase portrait (i.e. collection of solutions to (\ref{Liouville})) manifestly visible in the phase of  appropriately  modified   Liouville equation. The Gibbs equilibrium phase portrait will appear as an attractive fixed point in its phase portrait.

The modification of Eq.(\ref{Liouville}) that we search should ideally result from a thorough analysis if its solutions, in particular then from an analysis of  overall features of its solutions. We do not follow this path. Instead, we suggest a formal modification of (\ref{Liouville}) by adding to its right hand side
an appropriate (inspired by the Boltzmann collision term) particular realization of the second term on the right hand side of (\ref{generic1}).

When the macroscopic system is an ideal gas then, following Boltzmann, a separate treatment (separate from the free flow of particles) of binary collisions leads to such modification. For general macroscopic systems we replace binary collisions with transformations $z\leftrightarrow z'$ that preserve (i.e. $h(z)=h(z')$). We introduce thermodynamic force
\begin{equation}\label{GibbsX}
X(f^*)=f^*(z')-f^*(z)
\end{equation}
where $f^*(z)=S_{f(z)}(f)$ with S(f) given in (\ref{Gfund}).
With the dissipation potential
\begin{equation}\label{XiGibbs}
\Xi(f;z,z')=\int dz\int dz' W(f;z,z')\left(e^X+e^{-X}-2\right)
\end{equation}
the modified Liouville equation becomes
\begin{eqnarray}\label{modLiou}
\frac{\partial f}{\partial t}&=& L(f)E_f+\Xi_{f^*}|_{f^*=S_f}\nonumber \\
&=&-\frac{\partial}{\partial \rr}\left(f\frac{\partial h}{\partial \vv}\right)+\frac{\partial}{\partial \vv}\left(f\frac{\partial h}{\partial \rr}\right)+\int dz'W(f;z,z')(f(z')-f(z))
\end{eqnarray}
The function $W$ in (\ref{XiGibbs}) is required to satisfy the following three properties: (i) $W=0$ if $h(z)\neq h(z')$; (ii) $W>0$ if $h(z)=h(z')$; (iii) $W$ is symmetric with respect to
$z\leftrightarrow z'$.
Equation (\ref{modLiou}) (already suggested in \cite{GrmelaMT}) is indeed a particular realization of (\ref{generic1}). Also the  degeneracy  requirement (\ref{deg}) is clearly satisfied. Its  asymptotic solution is the equilibrium distribution function (\ref{caneq}) of the Gibbs theory. We note that the linearity of the Liouville equation has not been erased in its modification. The modified Liouville equation (\ref{modLiou}) remains a linear equation.

The dynamic formulation (\ref{modLiou})  of the static Gibbs theory  remains  formal.  Both the   physical  basis of the transformation   $z\rightarrow z'$ and  a detailed  analysis of solutions to (\ref{modLiou}) needed to complete the proof of the approach of its solutions to $f_{eq}(z)$ remain open. This type of investigation reaches beyond the scope of this paper.

\subsubsection{Roles of Energy and Entropy}

The roles that the energy and the entropy play in the Gibbs theory are the same as in the static Boltzmann theory. The total energy is an inner energy characterizing completely the individual nature of macroscopic systems. The entropy is universal and serves only to sweep away unimportant details (i.e. it plays only the role \textit{(Ent II)}). There is no internal  energy. As in the Boltzmann theory, there is no direct relation between the energy and the entropy on the starting level. Both are functions of  $f(z)$ but $f(z)$  cannot be eliminated between them.
The Boltzmann theory and the Gibbs theory differ  in the domain of applicability and in their dynamic formulations.  The static Boltzmann theory is applicable only to ideal gases, the static Gibbs theory to all macroscopic systems. In their dynamical formulations,  binary collisions that drive gases to equilibrium in the Boltzmann theory,  have no obvious parallel in the Gibbs theory. The transformations $z\rightarrow z'$ playing the role of binary collisions   in the dynamic extension of the Gibbs theory (\ref{modLiou})   remain formal.
The entropy in the Boltzmann theory is not postulated as in the Gibbs theory but emerges in the analysis of the time evolution. Independent arguments supporting the Gibbs entropy come for instance from the standard introduction of the Gibbs theory that we have recalled in the first paragraph of this section or from connections with the information theory \cite{Jaynes}.

As in the Boltzmann theory, there is no temperature on the starting level of the Gibbs theory because there is  no internal energy. The temperature arises in the submanifold of equilibrium states $f_{eq}(z)$ in the same way as in the Boltzmann theory.

\subsection{Euler-Navier-Stokes-Fourier fluid dynamics}\label{FM}

Starting levels in the two previous realizations of the mesoscopic thermodynamics  had no internal energy. Only after the passage to the target level  their  initial starting-level  inner energy turned to  the  target-level  internal energy on the level of equilibrium thermodynamics. In this section  we choose the level of fluid mechanics as the   starting  level. This level has both the inner energy and the internal energy.
The level of fluid mechanics  is an extension of the level of equilibrium thermodynamics to spatially inhomogeneous fluids. Fluids are composed of fluid particles. A fluid particle of  unit volume at  $\rr\in\mathbb{R}^3$ has the
momentum  $\uu(\rr)$,  an  internal structure characterized by the internal energy $\epsilon(\rr)$,  and the mass $\rho(\rr)$.
The  energy  $E$ that serves as the state variable on the level of equilibrium thermodynamics  turns in the extension  to the local internal energy $\epsilon(\rr)$, the number of moles $N$  to  $\frac{1}{M^{(mol)}}\rho(\rr)$, where $M^{(mol)}$ is the molar mass.
The volume $V$ on the level of equilibrium thermodynamics enters the extension in assigning   unit volume to fluid particles  and by requiring that the  energy, the mass and the entropy are   extensive  variables. The most important new feature in the extension is the motion of the fluid particles characterized by the new state variable $\uu(\rr)$. The motion of the fluid particles then brings  new contribution  $\frac{\uu^2(\rr)}{\rho(\rr)}$  to the energy (the kinetic energy) that is the inner energy since it is expressed in terms of the state variables.
The total energy is $E=E^{(in)}+E^{(int)}$, where $E^{(in)}=\int d\rr \frac{\uu^2}{2\rho}$ is the total kinetic energy and  $E^{(int)}=\int d\rr \epsilon(\rr)$ the total internal energy.

Global conservation $\frac{dA}{dt}=0$ of a quantity $A$ becomes  in the extension to the local field $a(\rr)$  (that is related to $A$ by $A=\int d\rr a(\rr)$)
the local conservation $\frac{\partial a(\rr)}{\partial t}=-\frac{\partial \JJ^{(a)}}{\partial \rr}$,  where  $\JJ^{(a)}(a(\rr))$ is a flux of the field $a(\rr)$. The local conservation of $A$ requires thus to specify  the flux $\JJ^{(a)}$ and boundary conditions. As to whether the local conservation implies the global one is decided by the boundary conditions.
In this paper we always assume  periodic boundary conditions that make integrals over boundary equal  zero. The local conservation thus always implies in this paper the global conservation.

Fluids that we investigate in this section on the level of fluid mechanics are characterized on the level of equilibrium thermodynamics by  fundamental thermodynamic relations (\ref{eqfund}).
Our objective is to make the passage  \textit{level of fluid mechanics}$\rightarrow$\textit{level of equilibrium thermodynamics} which ends up with a placement of the equilibrium thermodynamic state space inside the fluid mechanics state space (i.e. with  the fluid-mechanics equilibrium state) and the equilibrium fundamental thermodynamic relation  (\ref{eqfund}) with which the extension started.  In other words, we begin with fluids that are characterized on the level of equilibrium thermodynamics by (\ref{eqfund}), extend their investigation to fluid mechanics and then by following the  time evolution end up with the fundamental thermodynamic relation (\ref{eqfund}) with which the extension started.  Our focus is put again  on the roles that energy and entropy play in such passage.

The  variables
\begin{equation}\label{svfm}
x=(s(\rr),\rho(\rr),\uu(\rr))
\end{equation}
playing the role of state variables on the level of fluid mechanics are physically interpreted as   local entropy per unit volume, local mass per unit volume, and local momentum  respectively.

The fundamental thermodynamic relation is
\begin{eqnarray}\label{fhyd}
S(s,\rho,\uu)&=&\int d\rr s(\rr)\nonumber \\
E(s,\rho,\uu)&=& \int d\rr e(s,\rho,\uu;\rr)= E^{(in)}(\rho,\uu)+E^{(int)}(s,\rho,\uu)\nonumber \\
E^{(in)}(\rho,\uu)&=&\int d\rr \frac{\uu^2}{2\rho}\nonumber\\
E^{(int)}(s,\rho)&=&\int d\rr \epsilon(s,\rho;\rr)\nonumber \\
N(\rho)&=&\frac{1}{M_{mol}}\int d\rr \rho(\rr)
\end{eqnarray}
The energy $E^{(in)}(\rho,\uu)$ is the kinetic energy that is  in the classical hydrodynamics  the inner energy, i.e. the part of the energy that can be expressed in terms of state variables excluding the entropy. Moreover,  $s_{\epsilon}(\rr)=\frac{1}{\tau(\rr)}$  is physically interpreted as a local temperature and is assumed to be positive (an extension of (\ref{estar}) to local fields). If the function $\epsilon(s,\rho;\rr)$ is assumed to be the same as $E(S,N,V)|_{V=1, N=\frac{\rho}{M^{(mol)}}}$ on the level of equilibrium thermodynamics then such  assumption is called an assumption of local equilibrium. Identification of $\tau(\rr)$ with the local temperature is a weaker form of the local equilibrium assumption.
The positivity of the local temperature  implies that there is a one-to-one relation between the energy field $e(\rr)$ and the entropy field $s(\rr)$. For later use we recall the relations
\begin{equation}\label{relat}
s_e=\frac{1}{e_s};\,\,s_{\uu}=-\frac{e_{\uu}}{e_s};\,\,s_{\rho}=\frac{e_{\rho}}{e_s}
\end{equation}

With the fundamental thermodynamic relation (\ref{fhyd}) and the local equilibrium assumption,  the MaxEnt principle leads to the equilibrium state
\begin{equation}\label{hydeq}
\uu_{eq}(\rr)=0;\,\,\left(s_{e}(\rr)\right)_{eq}=\frac{1}{\tau_{eq}(\rr)}=\frac{1}{T};\,\,\left(s_{\rho}(\rr)\right)_{eq}=-\frac{\mu}{T}
\end{equation}
and  the fundamental thermodynamic relation  (\ref{yfre}). With this result we end  the static theory.

Now we turn to the dynamical theory.  We look for a particular realization of (\ref{generic1})  that replaces the postulated maximization in the MaxEnt principle.. We know already the potentials (\ref{fhyd}) entering (\ref{generic1}), we only need to identify the Poisson bracket expressing kinematics of (\ref{svfm}) and a dissipation potential.

Motion of continuum are transformations $\mathbb{R}^3\rightarrow\mathbb{R}^3$. These transformation form a Lie group. The momentum field $\uu$ is an element of the dual of the algebra corresponding to the group. The structure of the group manifests itself  in the dual of the  Lie algebra that corresponds to the group   as the Poisson bracket $\{A,B\}=\int d\rr u_i\left(\frac{\partial A_{u_i}}{\partial r_j} B_{u_j}-\frac{\partial B_{u_i}}{\partial r_j} A_{u_j}\right)$ \cite{Marsden},\cite{book} (the convention of  summation over  repeated indices is used). The remaining two fields $\rho(\rr)$ and $s(\rr)$ are let to be  passively advected by $\uu(\rr)$. The Poisson bracket that expresses kinematics of the state variables (\ref{svfm}) is  \cite{book}
\begin{eqnarray}\label{Pflm}
\{A,B\}&=&\int d\rr \left[u_i\left(\frac{\partial A_{u_i}}{\partial r_j} B_{u_j}-\frac{\partial B_{u_i}}{\partial r_j}\right)\right.\nonumber \\
&&\left.+\rho \left(\frac{\partial A_{\rho}}{\partial r_i}B_{u_i}-\frac{\partial B_{\rho}}{\partial r_i}A_{u_i}\right)\right.\nonumber \\
&&\left.+s \left(\frac{\partial A_{s}}{\partial r_i}B_{u_i}-\frac{\partial B_{s}}{\partial r_i}A_{u_i}\right)\right]
\end{eqnarray}
The Hamilton time evolution equation $\frac{\partial x}{\partial t}=L\mathfrak{E}_{x}$ with $x$ given in (\ref{svfm}) and $L$ in (\ref{Pflm}) are the Euler equations
\begin{eqnarray}\label{Euler}
\frac{\partial u_i}{\partial t}&=& -\frac{\partial (u_i\mathfrak{E}_{u_j})}{\partial r_j}-\frac{\partial p}{\partial r_i}\nonumber \\
\frac{\partial s}{\partial t}&=&-\frac{\partial (s \mathfrak{E}_{u_i})}{\partial r_i}\nonumber \\
\frac{\partial \rho}{\partial t}&=&-\frac{\partial (\rho\mathfrak{E}_{u_i})}{\partial r_i}
\end{eqnarray}
where
\begin{equation}\label{press}
p(\rr)=-e+\rho\mathfrak{E}_{\rho}+s\mathfrak{E}_s+u_i\mathfrak{E}_{u_i}
\end{equation}

We note in particular that the degeneracy requirements (\ref{deg}) hold and that the entropy contributes, through its presence in the local pressure $p(\rr)$, to drive the time reversible Hamiltonian time evolution. The gradient of the local pressure is partially an entropic force. Entropy takes this new role because of the presence of the internal energy. In the absence of such energy, as it is in the case of the Boltzmann and the Gibbs theories, the entropy does not participate in the Hamiltonian part of the time evolution.

The forces that drive fluids  to thermodynamic equilibrium are  the Fourier  force $X^F$ and the Navier Stokes forces $X^{NS}$ and $X^{NSvol}$
\begin{eqnarray}\label{NSF}
X_i^F(\rr)&=&\frac{\partial s_e}{\partial r_i}\nonumber \\
X_{ij}^{NS}(\rr)&=& \left(\frac{\partial e_{u_i}}{\partial r_j}+\frac{\partial e_{u_j}}{\partial r_i}\right)\nonumber \\
X^{NSvol}&=& div(e_{\uu})
\end{eqnarray}
Dissipation potential
\begin{equation}\label{Xifm}
\Xi^{flm}=\frac{1}{2}\int d\rr\lambda X_i^F(\rr)X_i^F(\rr)+\frac{1}{2}\int d\rr \eta X_{ij}^{NS}(\rr)X_{ij}^{NS}(\rr)+\frac{1}{2}\int d\rr \eta^{(vol)}(X^{NSvol})^2
\end{equation}
generates  the Navier Stokes Fourier gradient terms supplementing the Euler equations (\ref{Euler})
\begin{equation}\label{ENSF}
\frac{\partial}{\partial t}\left(\begin{array}{ccc}\uu\\e\\\rho\end{array}\right)=\left(\begin{array}{ccc}Euler\\equations\\ \end{array}\right)
+\left(\begin{array}{ccc}\Xi_{s_{\uu}}\\\Xi_{s_e}\\0\end{array}\right)
\end{equation}
The coefficients
$\lambda>0, \eta>0, \eta^{(vol)}>0$ introduced in the dissipation potential (\ref{Xifm}) are functions of (\ref{svfm}). The calculations that are  needed to turn (\ref{ENSF}) into the familiar Navier Stokes Fourier equations  use the relations (\ref{relat}) (see \cite{book} for details of the calculations)  .

\subsubsection{Extended fluid mechanics}

The level of kinetic theory playing the role of the starting level in Section \ref{B} involves more details than the level of fluid mechanics that plays the role of the starting level in this section. Indeed, the physical interpretation of the one particle distribution function $f(\rr,\vv)$ and of the hydrodynamic fields $(\uu(\rr),\rho(\rr))$  implies  that the hydrodynamic fields can be expressed in terms of $f(\rr,\vv)$ as its  moments $\rho(\rr)=\int d\vv f(\rr,\vv)$ and $\uu(\rr)=\int d\vv\,  \vv f(\rr,\vv)$. Are there autonomous levels that lie   between the level of kinetic theory and fluid mechanics? Such levels, if they exist,  are then expected to provide  a theoretical framework for  complex fluids (as for example viscoelastic polymeric fluids or suspensions) that are found to be outside the domain of applicability of the classical fluid mechanics.

Intermediate levels can be constructed in two ways: bottom up and top down. The former is an extension of the level of fluid mechanics by adopting extra fields (e.g. higher order  moments of $f(\rr,\vv)$ or fields characterizing an internal structure) as extra state variables. The latter as  a reduction from the level of the kinetic theory or other theories involving more details than fluid mechanics. We first discuss the former and then the latter constructions.

There are two kinds of the bottom up extensions.  The state variables in the first are  $x=(\uu(\rr),\rho(\rr),\aaa(\rr))=(\uu(\rr),\rho(\rr),a_1(\rr),...,a_k(\rr))$ and in the second $x=(\uu(\rr),\rho(\rr),s(\rr),\aaa(\rr))$.
In the first extension we  proceed as in the Boltzmann and the Gibbs theories  but with the fields $(\uu(\rr),\rho(\rr),\aaa(\rr))$ replacing the distribution function $f(\rr,\vv)$ (or $f(z)$ in the Gibbs theory). We need to specify the fundamental thermodynamic relation (i.e. we need to specify
$S=\int d\rr s(\uu(\rr),\rho(\rr),\aaa(\rr)),
E=\int d\rr e(\uu(\rr),\rho(\rr),\aaa(\rr)),
N=\int d\rr n(\uu(\rr),\rho(\rr),\aaa(\rr))$,
Poisson bracket expressing kinematics of the fields $(\uu(\rr),\rho(\rr),\aaa(\rr))$,  thermodynamic forces, and dissipation potential. As in the Boltzmann and the Gibbs theory,  there is no internal energy.
In the second extension we proceed as in fluid mechanics but with the kinetic energy replaced by an enlarged inner energy involving the extra fields $\aaa(\rr)$. The total energy involves   an internal energy that is expressed in terms of the entropy field that serves as one of the state variables.

The most familiar example of the top down extension is the reformulation of the Boltzmann kinetic equation (\ref{BEq}) into Grad's hierarchy \cite{Grad}. The one particle distribution function $f(\rr,\vv)$ is replaced by infinite number of fields $\aaa(\rr)^{(\infty)}(\rr)=(a_1(\rr),...)$ that are moments of $f(\rr,\vv)$ in the momentum $\vv$. There are two ways to rewrite (\ref{BEq}) into an equation governing the time evolution of $\aaa^{(\infty)}(\rr)$. First, it is a direct method \cite{Grad} consisting of multiplying (\ref{BEq})  by $v_i$ and integrating over $\vv$, then by $v_iv_j$ and integrating over $\vv$, and repeating this process for all moments. In the second method  we regard  (\ref{BEq}) as a particular realization of the GENERIC equation (\ref{generic}) and rewrite into moments separately all its building blocks (for example the Poisson bracket (\ref{idL}) is rewritten into  Grad's moments in \cite{GrGrad}). The two methods lead to two different infinite hierarchies \cite{GrGrad}.

If our objective is to obtain an extended fluid mechanics that involves a finite number of fields $\aaa(\rr)$, that  addresses physics of complex fluids,  and  that is compatible   with equilibrium thermodynamics in the sense of Section \ref{sec1} then, in both bottom up and top down approaches,   we need to identify the
finite number of fields $\aaa(\rr)=(a_1(\rr),...,a_k(\rr))$ joining the hydrodynamic fields $(\rho(\rr),\uu(\rr))$ and corresponding to them  building blocks of GENERIC. An inspiration can come from   an insight into the physics of the internal structure (e.g. various models of macromolecules composing polymeric fluids \cite{Kirkwood1}, \cite{Kirkwood2},  \cite{Bird}), from an insight into the geometrical structure of continuum dynamics (\cite{cosserat}, \cite{spain}, and from hierarchy reformulations of kinetic equations \cite{Grad},\cite{MR}, \cite{Rugg}, \cite{Jou}. Poisson brackets expressing kinematics of some of extended sets of hydrodynamic fields $(\uu(\rr),\rho(\rr),\aaa(\rr))$  are identified in \cite{Grmelacont},\cite{EssenPB}.

In the top down approach to extensions the passage from an infinite number of fields $\aaa^{(\infty)}(\rr)$ to a finite number  $\aaa(\rr)=(a_1(\rr),...,a_k(\rr))$ is called a closure of the infinite hierarchy. Difficulties encountered in finding  appropriate closures are well illustrated in the problem of finding Poisson brackets expressing kinematics of a finite number of moments $\aaa(\rr)=(a_1(\rr),...,a_k(\rr))$ in the Grad hierarchy. Two arguments developed in \cite{EssenTensors} suggest that there are only three autonomous fluid mechanics theories that can be based on the Grad hierarchy: $\mathbb{F}\mathbb{M}^{\infty}$ that is  fluid mechanics with all infinite number of moments playing the role of state variables (such fluid mechanics  is equivalent to kinetic theory), $\mathbb{F}\mathbb{M}^{5}$ that is the classical fluid mechanics with the classical hydrodynamic fields playing the role of state variables, and $\mathbb{F}\mathbb{M}^{\infty - 5}$ that is  fluid mechanics in which all Grad moments except the hydrodynamic fields play the role of state variables.
The first argument is based on the analysis of the Lie algebra of moments, the second argument is physical. We recall the latter argument. After the onset of turbulence the originally simple fluid whose behavior  is well described by the classical hydrodynamic fields becomes a complex fluid with an internal structure that needs higher Grad moments to describe its behavior. From the Kolmogorov cascade we know that   the complexity of the flow passes gradually to smaller  and smaller scales until the molecular scale is reached. At the molecular scale the
inner energy (i.e. the energy expressed in terms of $\aaa(\rr)$) of the complex turbulent flow turns into an internal energy.
If there were an autonomous fluid mechanics with a finite number $k$  of Grad moments then the Kolmogorov cascade would have a plateau. The turbulent decay would stop on the $k$-scale (i.e. the scale associated with $k$-th Grad moments), the flow would become $k$-laminar (i.e. laminar in the setting of the extended fluid mechanics in which $k$ Grad moments play the role of the state variables) and then a $k$-onset of turbulence would be needed to continue the turbulent decay to the molecular scale. No such plateau in the Kolmogorov cascade is, to the best of our knowledge,  observed.

\subsubsection{Roles of Energy and Entropy}\label{fmm}

Fluids are  composed of fluid particles that have  an internal structure and associated with it an internal energy. Its presence distinguishes the level of fluid mechanics from two levels discussed in the two previous illustrations of mesoscopic thermodynamics.
The total energy $e(\rr)$   of the fluid particle at the position $\rr$  is a sum of the inner energy (that is the kinetic energy  $\frac{\uu^2(\rr)}{2\rho(\rr)}$) and the internal energy $\epsilon(s,\rho;\rr)$. The total energy $E=\int d\rr e(\rr)$ of the fluid is then the sum of the energy of all fluid particles. The energy that generates the Hamiltonian part of the time evolution involves entropy and thus the entropy participates in the Hamiltonian time evolution. Specifically, the gradient of the local pressure $p(\rr)$ is the entropic force.
The entropy  generates  in addition also the dissipative forces driving the fluids to total equilibrium at which the fluid particles do not move and their internal structures are identical. The entropy plays thus on the starting level  both roles \textit{(Ent I)} and \textit{(Ent II)}. A weak local equilibrium assumption is carrying
the one-to-one relation between internal energy and entropy  from  the level of equilibrium thermodynamics to the level of fluid mechanics,    a complete local equilibrium assumption is carrying the complete equilibrium  thermodynamic relation  to individual fluid particles.

An interesting insight into the dependence of the passages \textit{starting level}$\rightarrow$ \textit{level of equilibrium thermodynamics}
on the starting level can be gained by comparing the passages with kinetic theory and fluid mechanics playing the role of starting levels. For all starting levels the time evolution making the passages are governed by (\ref{generic}). The differences are in the strength of the dissipation in the second term on the right hand side of (\ref{generic}).
In the context of fluid mechanics the passage from local equilibrium to the total thermodynamics equilibrium is made by the Navier-Stokes-Fourier dissipative forces (\ref{NSF}) and the dissipation potential (\ref{Xifm}). In the context of the Boltzmann kinetic theory the passage from  local Maxwellian distribution functions (that we can see as states corresponding to local equilibrium states in fluid mechanics) to the total Maxwellian distribution functions (\ref{ideq}) does not require an extra explicit dissipation. It is made simply by  coupling the dissipation driving  to the local Maxwellian distribution functions with the Hamiltonian completely nondissipative part of the time evolution (by the Grad-Villani-Desvillettes enhancement of dissipation \cite{Villani1}). More precisely, solutions of the Boltzmann equation get in the course of the time evolution only close to the local equilibrium and reach it only in the total equilibrium (\ref{ideq}). This comparison supports the conjecture that a very weak dissipation (instability) on the microscopic level is sufficient to grow by the Grad-Villani-Desvillettes enhancement of dissipation to macroscopic dissipation towards the level equilibrium thermodynamics.

Temperature on the level of fluid mechanics  has the same meaning as the temperature on the level of the equilibrium thermodynamics. The situation is different if the fluid mechanics is extended by adopting  fields $\aaa(\rr)$ as extra state variables. If the state variables of the extended fluid mechanics do not include the entropy field $s(\rr)$ then such setting is essentially the same as the setting of the Boltzmann and the Gibbs theory. There is no internal energy and no temperature on the starting level. If the entropy field $s(\rr)$ is included in the set of state variables then the energy that cannot be expressed in terms of $(\uu(\rr),\rho(\rr), s(\rr),\aaa(\rr))$ is the internal energy. But such internal energy is different from the internal energy in fluid mechanics in which the extra fields $\aaa(\rr)$ are missing in the set of state variables. The part of the energy $\widehat{\epsilon}(\aaa(\rr))$ that is expressed in terms of $\aaa(\rr)$ is excluded from the internal energy in the extended fluid mechanics. The meaning of  temperature depends on  what kind of walls we use in its measurement  and on what kind of control we have over the fields $\aaa(\rr)$. The walls can either pass or prevent passing the  energy $\epsilon(\rr)- \widehat{\epsilon}(\aaa(\rr))$ ($Walls^{(extfm)}$) or they are  the same as the ones used in the classical fluid mechanics in which the fields $\aaa(\rr)$ are not included in the set of state variables ($Walls^{(fm)}$). With $Walls^{(fm)}$ and no control over $\aaa(\rr)$ the measured  temperature is the same as in the classical fluid mechanics. With $Walls^{(extfm)}$ and with control over $\aaa(\rr)$ the measured temperature is a different temperature.

\subsection{van der Waals gas }\label{vdW}

In this section we extend the kinetic theory discussed in Section \ref{B}. The extension is not made by adopting extra distribution functions as state variables
(the one particle distribution function $f(\rr,\vv)$ remains the only state variable) but by  introducing  an internal energy. From the microscopic point of view, the van der Waals  (vdW) gas particles become interacting particles. Two types of interactions  are considered: a long range attraction and a short range hard-core  repulsion. The former is considered as an inner energy  and the latter as an internal energy. The inner energy is  the mean-field type Vlasov energy, the internal energy is expressed in a modification of the Boltzmann entropy. The hard core repulsion is replaced by  an excluded-volume type constraint.
From the point of view of hydrodynamic and equilibrium-thermodynamic type observations the vdW gas  differs from the ideal gas by experiencing transition from gas to liquid. The Vlasov gas shows the approach to spatially homogeneous distribution (Landau damping).  How do the roles of the energy and the entropy change when ideal gas becomes vdW or Vlasov gas?.

Before discussing the vdW gas on the level of kinetic theory, we recall the vdW  fundamental thermodynamic relation
\begin{equation}\label{vdWeqth}
S=\frac{5}{2}N+NR\ln \left[\left(\frac{E}{V}+a\frac{N}{V}\right)^{3/2}\left(\frac{V}{N}-b\right)\right]
\end{equation}
on the level of equilibrium thermodynamics
We note that (\ref{vdWeqth}) is a two parameter $(a,b)$ deformation of the ideal gas fundamental thermodynamic relation (\ref{eqideal}) on the level of equilibrium thermodynamics. If $a=0$ and $b=0$  then (\ref{vdWeqth}) reduces to (\ref{eqideal}). On the level of equilibrium thermodynamics, the relation (\ref{vdWeqth}) is based on heuristic arguments and experimental observations.

As in the previous sections we begin with the static theory. The state variable is the same one particle distribution function (\ref{svkt}) as in the Boltzmann theory.
The fundamental thermodynamic relation extending (\ref{idfund}) is \cite{vanKamp}
\begin{eqnarray}\label{vdWf}
S(f)&=&-\int d\rr\int d\vv(f(\rr,\vv)\ln f(\rr,\vv)+k_Bf(\rr,\vv)\ln(1-bn(\rr)))\nonumber \\
E(f)&=&\int d\rr \int d\vv \frac{\vv^2}{2}f(\rr,\vv)\nonumber \\
&&+\frac{1}{2}\int d\rr \int d\rr_1 \int d\vv \int d\vv_1  f(\rr,\vv)V_{pot}(|\rr-\rr_1|)f(\rr_1,\vv_1)\nonumber \\
N(f)&=&\int d\rr \int d\vv  f(\rr,\vv)
\end{eqnarray}
 where $n(\rr)=\int d\vv f(\rr,\vv)$.
 MaxEnt principle implies the equilibrium state
 \begin{equation}\label{eqvdW}
f_{eq}(\rr,\vv)=\left(\frac{1}{2\pi T}\right)^{3/2}n_{eq}(\rr) \exp\left(-\frac{\vv^2}{2T}\right)
 \end{equation}
 where $n_{eq}(\rr)$ is a solution to
 \begin{equation}\label{neqvdW}
 \ln n(\rr)+\frac{b}{1-bn(\rr)}+\frac{1}{T}\int d\rr'V_{pot}(|\rr-\rr'|)n(\rr')-\frac{3}{2}\ln (2\pi T)-\frac{\mu}{T}
 \end{equation}
 and the fundamental thermodynamic relation (\ref{vdWeqth}) on the level of equilibrium thermodynamics.

 Gas-liquid phase transition manifests itself in (\ref{eqvdW}) (as the familiar P-V-T curves  - see e.g  \cite{Callen}) and also in the manifold of equilibrium states (\ref{eqvdW}) (as a bifurcation in solutions to (\ref{neqvdW}) - see \cite{vanKamp}, \cite{Ireland}, \cite{book}).

Now we turn to the dynamic theory. Our objective again is to replace the  maximization of the entropy postulated in the static theory with the time evolution describing the experimentally observed approach to equilibrium.
How does the kinetic equation (\ref{BEq}) change when the ideal gas change into the vdW gas? There are at least two answers to this question. First it is a particular realization of (\ref{generic}) with the potentials (\ref{vdWf}) and the second  is the Boltzmann kinetic equation with two changes: the Vlasov term expressing the influence of the attractive force is added to its right hand side and the Boltzmann collisions of point particles are replaced by Enskog collisions of particles having a finite size. The modified  kinetic equation (called Enskog Vlasov equation \cite{GrmelaEV}) seems to be very natural from the physical point of view   but its compatibility with the static theory presented above as well as with the van der Waals fundamental thermodynamic relation (\ref{vdWeqth}) requires additional assumptions that do not appear to be natural from the physical point of view \cite{grm}, \cite{Ireland}. Below, we follow the path (already suggested in \cite{grmTRT}) on which the kinetic equation emerges as a particular realization of (\ref{generic}) with potentials (\ref{vdWf}).

The state variable is the same as in the Boltzmann kinetic theory, its physical interpretation is the same,  and thus also its kinematics expressed in the Poisson bracket (\ref{idL}) is the same. We leave also the same thermodynamic forces (\ref{XBol}) as well as  the thermodynamic potential (\ref{iddissp}). The binary collisions in which the entropy plays the role (\textit{(ENT II)} are thus left in the vdW gas the same as in the ideal gas. The finite size of the particles is taken into account in the Hamiltonian part of the time evolution by changing the available free energy $\mathfrak{E}(f)$ (\ref{vdWf}).  With the building blocks specified above, Eq.(\ref{generic}) becomes
\begin{equation}\label{vdWkeq}
\frac{\partial f(\rr,\vv)}{\partial t}=-\frac{\partial}{\partial \rr}\left(T f\frac{\frac{\partial\Phi}{\partial f}}{\partial \vv}\right) +\frac{\partial}{\partial \vv}\left(Tf\frac{\frac{\partial\Phi}{\partial f}}{\partial \rr}\right) -\frac{\partial\Xi}{\partial f^*(\rr)}|_{f^*(\rr)=\frac{\partial\Phi}{\partial f(\rr)}}
\end{equation}
Using the thermodynamic potential (\ref{vdWf}), the kinetic equation (\ref{vdWkeq}) gets the form
\begin{eqnarray}\label{Geq}
\frac{\partial f(\rr,\vv)}{\partial t}&=&-\frac{\partial}{\partial \rr}\left(\frac{\vv}{m}f(\rr,\vv)\right)\nonumber \\ &&+T\frac{\partial f(\rr,\vv)}{\partial \vv}\frac{\partial}{\partial \rr}\left(\ln (1-bn(\rr))-
\frac{bn(\rr)}{1-bn(\rr))}\right)\nonumber \\
&&+\frac{\partial f(\rr,\vv)}{\partial\vv}\frac{\partial}{\partial\rr}\int d\rr_1 V_{pot}(|\rr-\rr_1|)n(\rr_1)\nonumber \\
&&-\frac{\partial\Xi(f,f^*)}{\partial f^*(\rr,\vv)}|_{f^*=\frac{\partial\Phi}{\partial f}}
\end{eqnarray}
Solutions to this kinetic equation describe approach to the equilibrium vdW theory. Equation (\ref{Geq}) is a particular realization of the GENERIC equation (\ref{generic}). It  shares  the GENERIC structure with all well established equations describing approach to the level of equilibrium thermodynamics. The particularly interesting and new feature of (\ref{Geq}) is that
vdW equilibrium theory addresses gas-liquid phase transitions and consequently  (\ref{Geq}) addresses its dynamical aspects. In order to see them we need detailed solution to (\ref{Geq}). We hope to present them in a future paper.

\subsubsection{Roles of Energy and Entropy}

 When comparing the static kinetic theory of  ideal gases  (Section \ref{B}) with  the static kinetic theory of the vdW gas, the most  important new feature in the vdW theory
 is the presence of internal energy that is expressed as a modification (\ref{vdWf}) of the Bolktzmann entropy (\ref{idfund}).  From the physical point of view, the hard-core repulsive potential cannot be expressed in terms of only one particle distribution function (i.e. it is not an inner energy). At least two particle distribution function is needed to express it. In the setting of the kinetic theory in which only one particle distribution function serves as the state variable the hard core repulsive potential has to be considered as an internal energy. Its expression in terms of entropy then follows  from replacing the hard-core potential with the excluded-volume type constraint  \cite{vanKamp}.
We note that if we do not modify the entropy in (\ref{vdWf}) (i.e. if we keep only the Boltzmann entropy (\ref{idfund})) then there is no modification of the inner energy $E(f)$ in (\ref{vdWf}) that would imply the vdW fundamental thermodynamic relation (\ref{vdWeqth}) on the level of equilibrium thermodynamics. This  result is just another manifestation  of a well known result from equilibrium statistical mechanics, namely that the Gibbs equilibrium theory does not provide a setting for phase transitions unless it is somehow extended  (e.g. by carrying it to thermodynamic limit $N\rightarrow \infty;\,\,V\rightarrow \infty;\,\, \frac{N}{V}=const.$) \cite{Ruelle}. An internal energy that modifies the Gibbs entropy (in the setting of the Gibbs theory)  or  the Boltzmann entropy (in the setting of the Boltzmann theory) is needed  to make  phase transitions visible in  geometrical features  of  the manifold of equilibrium states  and  geometrical features of the fundamental equilibrium thermodynamics relation,   both obtained by MaxEnt.

In the dynamical view of the kinetic theory of the vdW gas we keep the setting of the static theory. As in fluid mechanics  (Section \ref{FM}), the internal energy (expressed in terms of the entropy in kinetic theory in (\ref{vdWf}))  generates one of the  forces that drive the reversible Hamiltonian mechanics. Unlike  fluid mechanics, the total entropy is not conserved. Only the available free energy (\ref{avE}) is. We note that if the extra term $f(\rr,\vv)\ln(1-bn(\rr)))$ in the entropy in (\ref{vdWf}) were  replaced by $f(\rr,\vv)\ln(1-bf(\rr,\vv)))$ then the modified entropy would remain  a Casimir of the Poisson bracket (\ref{idL}) and the total entropy would be conserved in the Hamiltonian part of the time evolution. With such modification the equilibrium states $f_{eq}(\rr,\vv)$ as well as the implied fundamental thermodynamic relation on the level equilibrium thermodynamics would be however modified. In particular, the dependence of $f_{eq}(\rr,\vv)$ on $\vv$ would be a modified Maxwell distribution. From the physical point of view (i.e. in order to guarantee the experimentally observed approach to equilibrium) the conservation of $\mathfrak{E}(f)$ and not the conservation of $S(f)$ is essential in the Hamiltonian part of the time evolution.

\section{\textit{starting level}$\longrightarrow$ \textit{target level with dynamics}}\label{Rate}

In this section we change the target level.  The level of equilibrium thermodynamics is replaced by a level involving less details than the starting level but still involving  the time evolution. Not only the roles of the energy and the entropy change but also the energy and the entropy themselves are  replaced by   different potentials.
The well known and  thoroughly investigated passage from kinetic theory to fluid mechanics will serve us as a guide.
There are essentially two points of view  of the passage \textit{ a mesoscopic level involving time evolution}$\longrightarrow$ \textit{ another  mesoscopic level that involves less details but still involves  time evolution}.
In the first we regard it as an intermediate step in the passage \textit{ a mesoscopic level involving time evolution}$\longrightarrow$ \textit{level of equilibrium thermodynamics}. In the second  we regard it in the space of vector fields rather than in the state space.

The first viewpoint is, roughly speaking, the viewpoint of Chapman and Enskog \cite{ChE}, \cite{GK} in their investigation  of \textit{level of kinetic theory}$\longrightarrow$ \textit{level of fluid mechanics}.  The idea is to look in the state space for an invariant (or approximately invariant) attractive manifold. The time evolution describing   approach to the invariant  manifold is then the time evolution describing the passage \textit{ a mesoscopic level involving time evolution}$\longrightarrow$ \textit{ another  mesoscopic level that involves less details but still involves  time evolution}. This viewpoint has two disadvantages. The first is that   passages to mesoscopic levels with time evolution
 exist also when the level of equilibrium thermodynamics is not well established (due to the presence of external forces) and consequently  the passage  \textit{ a mesoscopic level involving time evolution}$\longrightarrow$ \textit{level of equilibrium thermodynamics} does not exist.
 The second disadvantage is that the  thermodynamics  based on the approach to fixed points  (Section \ref{sec1}) does not directly extend  to thermodynamics based on  the approach to invariant (or approximately invariant) submanifolds.

On the other hand, when we lift the time evolution in the state space to the time evolution in the tangent (or alternative cotangent) space,  the approach involved in the passage \textit{ a mesoscopic level involving time evolution}$\longrightarrow$ \textit{ another  mesoscopic level that involves less details but still involves  time evolution} becomes the approach to fixed point (approach to vector fields governing the time evolution on the target level). The viewpoint of thermodynamics presented in Section \ref{sec1} becomes directly applicable. Moreover, this second viewpoint does not require that the approach to equilibrium states exists and thus is directly applicable to externally forced systems.
In the context of the extensive literature devoted to investigations of the passage \textit{level of kinetic theory}$\longrightarrow$ \textit{level of fluid mechanics} the second viewpoint of the approach to mesoscopic levels with the  time evolution
is essentially the one used in the Grad type investigations  \cite{Grad}.

We  follow below the second viewpoint. Our main objective is to show that the passage \textit{ a mesoscopic level involving time evolution}$\rightarrow$ \textit{ another  mesoscopic level that involves less details but still involves  time evolution} leads to thermodynamics but with  entropy and energy replaced by, roughly speaking, their rates. We shall  restrict our investigation in this section to working out one particular  example taken from chemical kinetics.

\subsection{Target level: Level of chemical kinetics (ChemKin)}

Chemical components $A_1,...,A_p$ undergo one chemical reaction
\begin{equation}\label{chr}
\alpha_1A_1+...+\alpha_pA_p\leftrightarrows \beta_1A_1+...+\beta_pA_p
\end{equation}
where $\alpha_1,...,\alpha_p$ and $\beta_1,...,\beta_p$ are stoichiometric coefficients. The state variables
\begin{equation}\label{svch}
x=\nn=\left(\begin{array}{ccc}n_1\\ \vdots\\n_p\end{array}\right)
\end{equation}
are number of moles of the components. Experimentally observed approach to chemical equilibrium $\nn_{eq}=\left(\begin{array}{ccc}(n_1)_{eq}\\ \vdots\\(n_p)_{eq}\end{array}\right)$ is governed by a particular realization of (\ref{generic})
\begin{equation}\label{rateXi}
\frac{dn_i}{dt}=-\Xi_{n_i^*}|_{\nn^*=\Phi_{\nn}}
\end{equation}
The potential  $\Phi(\nn)$ is a thermodynamic potential and $\Xi$ a dissipation potential.

We make a few observations about Eq.(\ref{rateXi}).

(i) If the dissipation potential $\Xi$ depends on $\nn^*$ only through its dependence  on the thermodynamic force  $X=\gamma_in^*_i$ (called in chemical kinetics chemical affinity) then (\ref{rateXi}) takes the form
\begin{equation}\label{ntev}
\frac{dn_i}{dt}=\gamma_iJ
\end{equation}
where
\begin{equation}\label{gamma}
\gamma_i=\alpha_i-\beta_i;\,\,i=1,...,p
\end{equation}
and
\begin{equation}\label{XJ}
J=\Xi_X
\end{equation}

(ii)
Particular forms of the potentials  $\Phi$ and $\Xi$ for which (\ref{rateXi}) becomes the familiar  Guldberg-Waage mass action law can be found in \cite{Grmchem}. In our analysis we can keep $\Phi$ and $\Xi$ unspecified.

(iii)  Since (\ref{rateXi}) is a particular realization of (\ref{generic}) its solutions
approach (as $t\rightarrow \infty$) equilibrium $\nn_{eq}$ that is a solution to $\Phi_{\nn}=0$.

\subsection{Starting level: Level of extended chemical kinetics (ExtChemKin)}

Chemical reactions, seen on the scale  of molecules,  are  complex processes described by quantum mechanics. Is there an intermediate level between the level on which the mass action law is formulated and the level on which the chemical reactions are seen on the scale of molecules? We follow here \cite{Grmchem} where the extension of (\ref{rateXi}) is made by including chemical inertia. The chemical flux $J$ introduced in (\ref{XJ}) is adopted as an extra state variable. It has been argued in \cite{Ajji} that this extension indeed carries the mass action law (\ref{rateXi}) towards the microscopic level. We assume that the extended mass action law described below represents a well established level.

With the state variables
\begin{equation}\label{svchJ}
x=\left(\begin{array}{cc}\nn\\J\end{array}\right)=\left(\begin{array}{cccc}n_1\\ \vdots\\n_p\\J\end{array}\right)
\end{equation}
the particular realization of (\ref{generic}) governing the time evolution takes the form
\begin{equation}\label{nJtev}
\left(\begin{array}{cc}\frac{d\nn}{dt}\\\frac{dJ}{dt}\end{array}\right)=\left(\begin{array}{cc}0&\Gamma\\-\Gamma^T&0\end{array}\right)
\left(\begin{array}{cc}\nn^*\\J^*\end{array}\right)-\left(\begin{array}{cc}0\\\Theta_{J^*}\end{array}\right)
\end{equation}
where $L=\left(\begin{array}{cc}0&\Gamma\\-\Gamma^T&0\end{array}\right)$ is a Poisson bivector  (see more in \cite{Ogul}),
\begin{eqnarray}\label{Gamma}
\Gamma&=&\left(\begin{array}{ccc}\gamma_1\\ \vdots\\ \gamma_p\end{array}\right)\nonumber\\
\Gamma^T&=&\left(\begin{array}{ccc}\gamma_1& \cdots & \gamma_p\end{array}\right)
\end{eqnarray}
are the stoichiometric matrices,
$\Theta$ is a dissipation potentiasl (its relation to the dissipation potential $\Xi$ appering in (\ref{rateXi}) is given later in (\ref{XiTheta})), and  $\nn^*=\Phi^{(ext)}_{\nn}$,
$J^*=\Phi^{(ext)}_{J}$, where $\Phi^{(ext)}(\nn,J)$ is a thermodynamic potential extending the thermodynamic potential $\Phi(\nn)$ introduced in (\ref{rateXi}).

Regarding solutions to (\ref{nJtev}), we note  that this equation is a particular realization of (\ref{generic}) and thus the thermodynamic potential $\Phi^{(ext)}(\nn,J)$ plays the role of the Lyapunov function indicating
approach to equilibrium state $\left(\begin{array}{cc}\nn\\J\end{array}\right)_{eq}$ that is a solution to $\Phi^{(ext)}_{\nn}=0,\,\,\Phi^{(ext)}_{J}=0$. Static view of (\ref{nJtev}) is the MaxEnt passage to the equilibrium state $\left(\begin{array}{cc}\nn\\J\end{array}\right)_{eq}$.
Other properties of solutions to Eq.(\ref{nJtev}) that address its  relation  to (\ref{rateXi}) are discussed in the next section.

\subsection{(ExtChemKin)$\longrightarrow$ (ChemKin)}

Now we are in position to investigate in the setting of chemical kinetics described above the passage  \textit{ a mesoscopic level involving time evolution}$\longrightarrow$ \textit{ another  mesoscopic level that involves less details but still involves  time evolution}. The target level is the level represented by (\ref{nJtev}) and the starting  level is represented by (\ref{rateXi}).

If the term $\Theta_{J^*}$ is dominant in the second equation (\ref{nJtev}) then $J$ evolves faster that $\nn$. As a first approximation, we can solve (\ref{nJtev})  by following first the time evolution governed by the second equation in (\ref{nJtev}) to its conclusion and then follow the time evolution governed by the first equation in (\ref{nJtev}) in which $J^*$ is replaced by $J^*_{stat}(\nn^*)$  that
is a solution to
\begin{equation}\label{Psistat}
\Psi_{J^*}(J^*,(J^*)^{\dag})|_{(J^*)^{\dag}=-<\Gamma^T,\nn^*>}=0
\end{equation}
where
\begin{equation}\label{Psi}
\Psi(\nn,J^*;(J^*)^{\dag})=-\Theta(\nn,J^*)+J^*(J^*)^{\dag}
\end{equation}
Solution of (\ref{Psistat}) is
\begin{equation}\label{Jstssol}
J^*_{stat}(\nn^*)=\Theta^{\dag}_{(J^*)^{\dag}}((J^*)^{\dag})|_{(J^*)^{\dag}=-<\Gamma^T,\nn^*>}
\end{equation}
where $\Theta^{\dag}((J^*)^{\dag})$ is the Legendre transformation of $\Theta(J^*)$.

After $J$ reached the state at which $J^*=J_{stat}(\nn^*)$ the time evolution continues as the time evolution of $\nn$ governed by the first equation in (\ref{nJtev}) in which  $J^*$ is replaced by $J^*_{stat}(\nn^*)$ given in (\ref{Jstssol}).
The time evolution of $\nn$ is thus governed by
\begin{equation}\label{1Theta}
\frac{dn_i}{dt}=-\Theta^{\dag}_{\nn^*}((J^*)^{\dag})|_{(J^*)^{\dag}=-<\Gamma^T,\nn^*>}
\end{equation}
By comparing this equation with Eq.(\ref{rateXi}) we see that the two dissipation potentials $\Xi$ and $\Theta$ are related by
\begin{equation}\label{XiTheta}
\Xi(\nn,\nn^*)=\Theta^{\dag}((J^*)^{\dag})|_{(J^*)^{\dag}=-<\Gamma^T,\nn^*>}
\end{equation}

Summing up, we have reformulated (\ref{nJtev}) into  two equations:  Eq.(\ref{rateXi}) and
\begin{equation}\label{1stage}
\frac{\partial J^*}{\partial t}=\mathbb{G}\Psi_{(J^*)^{\dag}}|_{(J^*)^{\dag}=-<\Gamma^T,\nn^*>}
\end{equation}
In (\ref{1stage}) we have used $J^*=\mathbb{G}J$, where  $\mathbb{G}=\Phi^{(ext)}_{JJ}$,  and  (\ref{XiTheta}) in (\ref{rateXi}).  The three equations (\ref{nJtev}), (\ref{rateXi}) and (\ref{1stage}) describe  three reductions. We emphasize again that the existence of all three reductions is guaranteed by our assumption that
all three levels (i.e. the level of equilibrium thermodynamics, the level of the Guldberg-Waage chemical kinetics, and the level of the extended Guldberg-Waage chemical kinetics) are  autonomous well established levels.
All tree equations (\ref{nJtev}), (\ref{rateXi}), and (\ref{1stage}) are particular realization of (\ref{generic}). The first two  describe approach to the level of equilibrium thermodynamics and the third  approach to the level of the Guldberg-Waage chemical kinetics. The first two equations are thus  two additional  examples of the time evolution equations investigated in Section \ref{sec1}, the third is new. We shall now discuss it in a more detail.

First,  we establish terminology. In order to make a clear distinction between   passages \textit{starting level}$\rightarrow$\textit{level of equilibrium thermodynamics} and \textit{starting level}$\rightarrow$\textit{target level}  with  target levels that are different from the level of equilibrium thermodynamics we use the adjective "rate" in investigations of the latter.  We call the thermodynamic potential $\Psi$ in (\ref{Psi}) a rate thermodynamic potential, $\Theta$ in (\ref{1stage}) a rate entropy, the time evolution governed by (\ref{1stage}) a rate time evolution. The adjective "rate" points to the fact that the space in which the time evolution takes place is the tangent (or alternatively cotangent) space.

Both the entropy  $S^{(ext)}(\nn.J)$ and the rate entropy  $\Theta(\nn,J^*)$ appear in both equations (\ref{nJtev}) and (\ref{1stage}). Their roles are however very different. In  Eq.(\ref{nJtev}) (governing the reduction to  the level of equilibrium thermodynamics)  the entropy $S^{(ext)}(\nn.J)$ drives the reduction and the rate entropy $\Theta(\nn,J^*)$ plays the role of the geometrical structure transforming gradient of the entropy into a vector. On the contrary,  in Eq.(\ref{1stage}) (governing the reduction to the level of the Guldberg-Waage chemical kinetics) the rate entropy $\Theta(\nn,J^*)$ drives the reduction and the entropy $S^{(ext)}(\nn.J)$ (in the form of the Hessian
$\mathbb{G}=\Phi^{(ext)}_{JJ}$) is the  geometrical structure in the vector field.

In a small vicinity of thermodynamic equilibrium states
the rate entropy $\Theta$ can be seen, roughly speaking, as entropy production in the Guldberg-Waage time evolution. This follows from:
(i) the relation (\ref{XiTheta}) between the dissipation potential $\Xi$ and the rate entropy, (ii) from $\frac{d\Phi}{dt}=-<n^*,\Xi_{n^*}>|_{\nn^*=\Phi_{\nn}}$ implied by (\ref{rateXi}), from the fact that  close to equilibrium all dissipation potentials are quadratic (see the text following (\ref{Xi})), (iii) and from the fact that Legendre transformations of quadratic potentials as well as the scalar product of the variable with   gradient of the potential remain quadratic potentials..

Skeptical views of the relevance of  the entropy production in mesoscopic dynamics, occasionally  appearing in the literature,  originate  most likely  from unsuccessful attempts to see the entropy production  as Lyapunov-like function in the time evolution equations describing approach to thermodynamic equilibrium states. The entropy production (more precisely the rate entropy) does play such role but in different time evolution equations. Equations describing the  approach to equilibrium stats are  replaced by equations describing the approach to lower level  (i.e. involving less details than the dynamics on the starting level)   mesoscopic dynamics.
The rate entropy has moreover a more general applicability than the entropy since the approach to a lower level dynamics exists also when the approach to thermodynamic equilibrium states does not exist (e.g. due to presence of  external forces).

Finally, we make an observation about the static version of (\ref{1stage}). We recall that the  static version of (\ref{rateXi}) is the MaxEnt principle. The passage from the starting level to the level of equilibrium thermodynamics, made by following the time evolution governed by  (\ref{rateXi}),  is made in the  static version of (\ref{1stage}) simply by maximizing the  entropy subjected to constraints $(\Phi_{\nn}=0)$. Analogically, the static version of (\ref{1stage}) is minimization of the rate entropy subjected to constraints (i.e. by solving $\Psi_{J^*}=0)$. This view of (\ref{1stage}) is in fact  the Onsager variational principle \cite{Ray}, \cite{OnP}, \cite{OM}, \cite{Doi} applied to  chemical kinetics. The time evolution equation (\ref{1stage}) can be seen thus as an extension of the Onsager variational principle to chemical kinetics  similarly as (\ref{rateXi}) is an extension of MaxEnt to dynamics. When seeing (\ref{1stage}) in the context of the Onsager variational principle, the rate thermodynamic potential $\Psi$ is called Rayleighian. We note that the rate entropy $\Theta$ decreases to its minimum in the time evolution governed by (\ref{1stage}) . We can therefore call the minimization of the rate entropy MinRent similarly as we call the maximization of the entropy MaxEnt.

\section{Concluding Remarks}

The experimentally observed approach to equilibrium in externally unforced macroscopic systems is driven by gradients of energy and entropy. If the complete energy can be expressed in terms of the chosen state variables then the gradient of energy generates the reversible Hamiltonian time evolution and the gradient of entropy the  irreversible time evolution during which unimportant details are swept away and  the equilibrium pattern in the phase portrait emerges. This situation is illustrated in the Boltzmann and the Gibbs theories.
If there is a part of energy  that  cannot be expressed in terms of the chosen state variables (called an internal energy) then the  entropy is used to express it and the gradient of entropy becomes also one of  driving forces of the reversible Hamiltonian time evolution. This situation is illustrated in fluid mechanics and kinetic theory of the van der Waals gas. The latter illustration also shows that the gas-liquid phase transition appears  as a geometrical feature of the manifold of equilibrium states and of the equilibrium fundamental thermodynamic relation implied by the kinetic theory  only if the energy in the kinetic theory formulation involves an internal energy (in the van der Waals gas it is the energy generating hard-core repulsive forces) that cannot be expressed in terms of one particle distribution function and is expressed in terms of entropy.

The experimentally observed approach  to mesoscopic dynamical theories involving less details is driven by gradients of potentials that are related to rates of the energy and the entropy. This situations is illustrated in chemical kinetics. The static version of such  reduction becomes  the Onsager variational principle in the setting of chemical kinetics.

\textbf{Acknowledgement}
\\

I would like to thank O\v{g}ul Esen, V\'{a}clav Klika, and Michal Pavelka,  for stimulating discussions.
\\

\end{document}